# Empowered and Embedded: Ethics and Agile Processes

Niina Zuber (niina.zuber@bidt.digital)[1], Severin Kacianka (severin.kacianka@tum.de)[2], Jan Gogoll (jan.gogoll@bidt.digital)[1], Alexander Pretschner (alexander.pretschner@tum.de)[1,2], and Julian Nida-Rümelin (julian.nida-ruemelin@lrz.uni-muenchen.de)[1,3]

Abstract:

In this article we focus on the structural aspects of the development of ethical software, and argue that ethical considerations need to be embedded into the (agile) software development process. In fact, we claim that agile processes of software development lend themselves specifically well for this endeavour. First, we contend that ethical evaluations need to go beyond the use of software products and include an evaluation of the software itself. This implies that software engineers influence peoples' lives through the features of their designed products. Embedded values are thus approached best by software engineers themselves. Therefore, we put emphasis on the possibility to implement ethical deliberations in already existing and well established agile software development processes. Our approach relies on software engineers making their own judgments throughout the entire development process to ensure that technical features and ethical evaluation can be addressed adequately to transport and foster desirable values and norms. We argue that agile software development processes may help the implementation of ethical deliberation for five reasons: 1) agile methods are widely spread, 2) their emphasis on flat hierarchies promotes independent thinking, 3) their reliance on existing team structures serve as an incubator for deliberation, 4) agile development enhances object-focused techno-ethical realism, and, finally, 5) agile structures provide a salient endpoint to deliberation.

Software might not literally be "eating the world" (Andreessen 2011) – it did, however, revolutionize the way we do business, then changed the way we live and has now begun to influence the way we think. While software is supposed to enhance human abilities, it seems, sometimes, that it rather reduces the number of options that might be available to a person. The development processes for software were mainly designed as a response to the requirement for software to adapt to business needs which in turn change rapidly in a globally connected world. Software can influence decisions, amplify bias, exacerbate unfairness and perpetuate existing inequalities and it should therefore no longer be reduced to a mere tool for doing business. Software developers need to consider the implications of their products which in turn depend on the values (implicitly) embedded into the software design. When weighing two options, for example, developers cannot solely focus on technical and business needs: They need an *ethos* to moor their decisions and a *prâxis* that steers sound ethical deliberation.

In this paper, we focus on *prâxis* that entails types of actions that constitute life forms instead of individual actions and that by simple addition makeup lifeplans (c.f. Nida-Rümelin 2020). Hence, we focus on structural aspects of the development of ethical software, and argue that ethical considerations need to be embedded into the software development process. Agile development processes naturally lend themselves to ethical considerations because they empower developers to work in small teams and solve problems independently, without tight central control (c.f. Spreitzer 2008, Tariq et al. 2016). This approach gives developers substantial leeway in influencing the design of the system. However, agile software development is no silver bullet. There are many situations in which agile methods are more difficult to implement. Examples include highly regulated industries including medical, aerospace, and automotive, where regulation often demands specific processes and elaborate documentation; the design of physical products where specifications often need to be decided a long time before software development begins; large-scale projects; or corporate cultures that are highly hierarchical (e.g., Poth et al. 2020). Agile development practices work particularly well in contexts where a small team of developers builds a product and requirements for said product are subject to frequent change. Here, the iterative nature of agile processes and the close communication with the customer improves the end result, because customers get to see early interim results and can thus help clarify their requirements. To oversimplify, true agile practices are more often found in teams that develop websites or phone apps than companies that develop braking systems or EEG monitors. Traditional software development often incorporates a reduced version of ethical deliberation in the form of a technology assessment or dedicated safety or security analysis (c.f. Schneier 1999; Ruijters and Stoelinga, 2015; Grunwald 2018). While many technology assessments concern risk forecasts, it is only since the early 2000 that ethical technology assessments came into play such as the VDI 3780 standard (Verein Deutscher Ingenieure 2000) or Palm and Hansson's eTA approach (Palm and Hansson 2006). In regard to the ethical design of software systems the IEEE 7000/ D3 (2020) focuses on team members' competencies that are necessary in order to consider and deal with ethical issues while developing software. In doing so, they stress that ethical theories help in evaluating software systems by identifying ethical values prologing to their consequences, to practices, or in aspects of universalist claims (IEEE 7000/ D3 2020, pp. 13-22). Hence, to permanently include an ethicist in development teams is suggested by McLennan et al. (2020). However, this seems unattainable as it would be prohibitively expensive for smaller companies as well as the fact that there exists a general shortage of people with these skills. Therefore, we will argue that a systematic ethical deliberation is particularly suited to be embedded into agile processes without the permanent need for trained ethicists.

---

[1] Bavarian Research Institute for Digital Transformation
[2] Technical University of Munich
[3] Ludwig Maximilian University of Munich

Introducing any approach to implementing ethical deliberation to software development that would require practitioners to completely change their way of working would be an uphill battle. Fundamental changes to a management culture are costly (both in terms of money and time) and require strong commitments from developers, designers, and management (Gablas et al. 2018). Ethics certainly is an important factor to consider in software development, but it is by far not the only one. Software engineers and designers want to build products; and companies have to create cash flows. Therefore, instead of creating an entirely new way of doing business, we suggest including ethics into a working style that is already widely used and has features and capabilities that promote and foster ethical deliberation. This approach is closely linked to bottom-up ethical approaches that refrain from Principlism in ethics (Dancy 2004; 2018) - a dominant trend in modern times since the rise of European rationalism.

We put forward the claim that, especially in the domain of software development, agile methods (Scrum, extreme Programming, Essence etc.) offer many features that enable the development of ethically informed products throughout the entire development process. If ethical deliberation is to become a part of software development culture, it must be compatible with the processes used by software developers, designers, and operators; any other approach bears the risk of ethics to be ignored or reduced to a mere window-dressing effort. Ethical deliberation of information technology (IT) requires in-depth knowledge of the technical feasibility of normative criteria. However, even if the required expertise existed, it has, so far, not been integrated into an entrepreneurial culture.

# 1 Empower the engineer: autonomy as a requirement for good normative design

Morally illegitimate IT entails software products that might, often accidentally, exclude people due to their capabilities; make people change their daily life without their consent; display information consent forms in such a way that the user feels overwhelmed; or nudge people to spend hours on their homepages or apps (cf. Harris 2016; Coeckelbergh 2021; Mathur et al. 2019; Mathur et al. 2021). Conversely, morally desirable software systems include aspects of integration, minimization of discrimination, reduction of sexism and racism, fostering mutual respect, inclusivity, trust, and taking into consideration social structures, sustainability issues and moral reasons (cf. Nissenbaum 1999; Friedman et. al. 2008; Zwart 2014; Nida-Rümelin and Weidenfeld 2018; Steed and Caliskan 2021; van Wynsberghe 2021).

Morally desirable products are difficult to identify and test, since it is impossible to define decisive criteria that have to be met in exactly the same way in all possible scenarios. Ethical deliberations cannot be realized by using checklists or with the help of predefined answers (cf. Gogoll et al. 2021). Therefore, it remains essential to rethink the entire process for each new design project as well as product deployment and maintenance. In computer science, this is called the software development life cycle (Kneuper 2017). Hence, we need to identify principles or values that are relevant in that particular case. In doing so, we need to exercise judgmental reasoning (cf. Rohbeck 1993) to know how to apply general principles or values to particular cases and within the whole life cycle. The relationship between principles and values and their application in particular circumstances cannot be grasped by deduction or subsumption only. It is rather a rational-hermeneutic deliberation that cannot be understood in any algorithmic pattern.

Values are often regarded as the key starting point to enable normative deliberation in software engineering (cf. Spiekermann 2015; Senges et al. 2017; Spiekermann and Winkler 2020). This reasoning also inspires Codes of Conduct and Codes of Ethics as a means of establishing certain values that either shall not be violated by software or shall be taken into account in development. Values that are mentioned in the context of information technologies are, for instance: transparency, privacy, freedom of bias (see Friedman and Hendry 2019), accountability, fairness, cybersecurity, or sustainability. Those values open up a space for normative thinking (cf. Gogoll et al. 2021): They represent a starting point to think about what we ought to do in this particular case in terms of that desirable value: Is privacy an issue for this software product? How is privacy best understood in this case: Do the right people have access to the data or is the data protected against access by third parties (security)? The challenge, then, is how to interpret values and translate them into technical solutions. Codes of Conduct are certainly useful for establishing a set of (potential) values. For the development of ethical software, however, they do not constitute more but a mere starting point. Codes of Conduct typically comprise a set of higher-level regulative ideas, whose status is more or less self-evident, but they do not set forth a method of how to weigh these different ideas and to solve potential conflicts between them. They are to a large extent neutral regarding different theories of normative ethics, which is an advantage in terms of acceptability, but a disadvantage in terms of practicability. Hence, they cannot be more than a starting point, especially when we consider that software is deployed in almost all contexts today. Not surprisingly, the characteristics of software also turn out to be very context-specific (Briand 2017), as does the practice of software engineering. It therefore seems utterly unrealistic to expect a one-size-fits-all solution, i.e., a solution applicable to any kind of software, for embedding ethical values into software or applying ethical rules to the development process . The fact alone that software is context-specific requires a customized approach for each new software product. Values are ideals to which people aspire. There are different types of values, economic values are motivated by economic reasons such as efficiency and profit, political values by political reasons and so forth. Moral values are motivated by moral reasons which means that their motivation goes beyond any personal interest an agent may have. Moral reasons often vary across cultures and can change over time. They can, however, be transformed into universal laws or incorporated into morally desirable behavior, i.e., virtues. To stress this point: A value such as *care* is displayed in a concrete principle such as "do no harm to others". This can be formulated as a right once it has been recognized as a legitimate claim, such as the Offences against the Persons Act. Alternatively, it can be expressed in a moral norm which commands that you shall not abuse anybody, neither physically nor psychologically. But how can we preserve *normative orientation knowledge* (cf. Nida-Rümelin 2016) while reasoning technologically, i.e., using technological languages and concepts? In other words, in what ways do IT devices unsettle and challenge knowledge about normatively desirable behavior in concrete situations? If their normative impact is to be considered, digital artefacts cannot be explained in terms of natural science only, i.e., they cannot be exclusively described and explained in terms of technological functionality. They influence our mutual expectations and demands. This in turn, is precisely why some digital artefacts transform entire societal areas. Understanding this transformation is also a conceptual, analytic and hermeneutic undertaking that comes in addition with technical know-how (cf. Spiekermann 2016, Friedman and Hendry 2019, Reijers and Coeckelbergh 2020, Zuber et al. 2020). The digital product must be thought of as structuring social contexts, human capabilities, and transforming the way humans perceive or understand the world (cf. Ihde 1996).

It seems reasonable to address values in IT within the framework of virtue ethics, i.e., to think about whether a particular information technology fosters or undermines e.g., moral, sustainable or autonomy-respecting practices. In virtue ethics moral practice is defined as a collective action whose purpose is shared and designated as valuable, e.g., honesty, friendship or care (cf. MacIntyre 1981). If IT hinders people from establishing and nurturing caring relationships and even brings about behavior that makes it easier to stalk, bully, humiliate, or defame people, it is certainly worthwhile to rethink its design or – in the worst case – think about whether to deploy it at all. Virtue ethics highlights the fact that people who show the trait of being a caring person will act accordingly without constantly having to cognitively deliberate whether they want to be a caring person.

This makes virtue ethics a viable approach in the domain of software systems, in which desirable behaviors, in the classical sense *virtues*, are applied to techno contexts (cf. Vallor 2016; Reijers and Coeckelbergh 2020). Once desirable attitudes are formulated, they need to be promoted by practices. Vallor (2016) focuses on twelve techno-moral virtues: 1. honesty, 2. self-control, 3. humility, 4. justice, 5. courage, 6. empathy, 7. care, 8. civility, 9. flexibility, 10. perspective, 11. magnanimity, 12. techno-moral wisdom. In contrast to values applicable to a technical design, virtues pertain to the characters of the developer as well as to the user or other stakeholders. The habituation of those twelve good techno virtues will promote a good life and reduce moral wrongdoing through the deployment and use of technological devices. Thus, Vallor is not exclusively addressing software developers, but the entire community where IT is deployed. Therefore, desirable virtues need to be taken into account in the design of a product insofar as desirable practices should not be undermined by the development, deployment, and use of IT. Virtue ethics are necessary for the development of software, since it is important that software developers should develop those virtues themselves. This, in turn, means that management and development cultures should encourage and foster techno-virtuous behavior by introducing desirable practices into the development process. This also includes considering users and other indirect stakeholders whose life will be affected by the system. Therefore, it is important to develop a product in such a way that it will not undermine morally good user behavior, but instead support desirable outcomes. This means that any deliberation, decision-making and action within the context of software development as well as handling the product throughout the entire lifecycle must be in line with techno virtues.

It requires practical reasoning to explicate and navigate the context between values, goals and specific requirements that guide implementation (for an overview of goals in requirements engineering, see van Lamsweerde (2001)). This skill should be learned and practiced, esp. because normative competence is not only an epistemic endeavor, which would mean that the main goal is an increase in knowledge, but rather a technique: It is exercising judgment, or in Vallor's terminology, disposing of *technomoral wisdom*. It must be emphasized that this power of judgment is also an attitude. This enables the individual to acquire an ethical awareness about the necessity for normative deliberation if the developed products are to be moral, sustainable, or promoting democractic values etc. This means, then, that practical reasoning is not a purely cognitive process, but also a *hexis:* a stance towards the world. We understand virtues (*arete*) rather in Aristotle's classic sense that a virtue (*arete*) is not only a result of habituation (*ethos*), but also entails a krisis (*decision*) and a hexis (*attitude*) (c.f. Nida-Rümelin 2020). In modern virtue ethics this threefold constitution of virtues is often left out and virtues are reduced to result from social habituation only (c.f. McIntrye 1981). This leads to the question: who is in a position to deal with normative issues that are caused by or relate to technical devices? Obviously, ethical issues need to also be addressed from the engineers' perspective. We need to foster practical reasoning with regard to software engineering. On the one hand, this is a question of curricula and education (cf. Shen et al. 2021; Anderson (ed.) 2017; Tavani 2013, esp. Chap. 3). On the other hand, it is relevant for enabling or encouraging normative reasoning within the working environment. Ethical deliberation requires sufficiently effective structural integration, precisely because of the contingency that is entailed by pluralism of reasons. Hence, we have again established the context-sensitivity of ethical deliberations in regard to information technology: Some deliberations are rather straight-forward, while others remain more difficult to detangle. Still, integrating ethical deliberations into development processes is an essential step when it comes to developing ethically desirable software.

## 2 Why is software so special?

If the software engineer is expected to deliberate on moral issues and not only to focus on technical issues in terms of feasibility and functionality, digital artifacts should be understood as purporting or implying values itself. However, this is not to be misunderstood as the digital product creating its own moral values. What it means is that the digital artifact itself incorporates the value attitudes of the designer. This is called the *embedded values approach* (cf. Friedman and Nissenbaum 1997; Brey 1998, 2000; Van Wynsberghe and Moura 2013). In John Moor's (2005) terminology: the subjects of this analysis are *ethical impact agents* and *ethical implicit agents*, that is, those software systems that either cause external negative moral effects or those in which (distorted) moral expressions have already been implemented. Accordingly, the analysis excludes attempts to develop *explicit and completely ethical agents*, i.e., artificial agents that are supposed to be able to autonomously perform deontic deliberations in a certain domain or across many situations, i.e., agents that can deliberate morally and justify their decision in a well-founded way.

To be able to adopt the embedded values approach, we have to abandon the idea of the means-character of information technology that comes with the traditional thesis of value neutrality of technology. The means-purpose nature of technology, which is inherent in tools, does not, however, directly apply to information technology, because information technology comprises preset products over which the user has no complete control (cf. Hubig 2015; Grunwald 2015): Take washing machines, search engines or assistance systems in robotics. They already contain many assumptions about the water consumption, the power consumption or about the placement procedures of the search results up to the external design of a robot, which can be anthropomorphic or zoomorphic and thus simulate a lively encounter. All these features are not neutral insofar as they contain or convey values. The way in which the user can interact with this kind of digital technology is already partially predetermined by the technology itself which means that the user's autonomy is restricted in the interaction. B.J. Fogg (2002) refers to this characteristic of many IT-systems as *persuasiveness*. This concept may be partially transferable to classical technology insofar as every technology requires certain behaviors, because certain reactions are usually necessary for technology to be functional. In the case of information technology, however, many pre-formed properties and infrastructures already have certain values which are not immediately obvious, and which cannot be influenced from the outside (or only to a limited extent). Additionally, information technology is implemented throughout many domains of daily life and its determining force is often quite invisible, impenetrable, and multidimensional and thus difficult to grasp. Those phenomena are subsumed under the concept of *opacity* (cf. Brey 2000; van den Eede 2010). It is claimed that information technology is of an opaquer nature than many traditional tools such as doors, hair dryers or hammers. Information technology is not restricted to one specific working context, precisely because it can serve a variety of purposes. Just think of the telephone receiver that one puts down so that one is not reachable at a given moment. Surely that is not the intended purpose of the telephone receiver. This surplus of purposes (cf. Rohbeck 1993) makes an ethically defined scenario difficult sometimes, since the use and orientation can change at any time (cf. Vallor 2016). Even the speed with which software products are developed makes them rather unique. Moreover, executing *mind work* (Agar 2019), information technology substitutes or supports not only a pure physical force but is often placed within the very core of humanity as such: It strikes into the marrow of sociality, emotionality and reason. This can be illustrated by information technology supporting communication (e.g., social media or video calls) or decision-supportive analysis and recommendation software assisting medical diagnosis, application processes or parole conditions. Many information technologies transform the way we live precisely because they impact that which is genuinely human: how we express our emotions, how we judge, how we evaluate trust relationships, and much more. Thus, information technology can realign reciprocal normative commitments that sustain our sociality (cf. Nida-Rümelin 2019). These are all very compelling reasons for demanding and imposing ethical deliberation, but also for why this endeavor is difficult and complex.

In the case of information technology, it remains unclear in many cases which questions are the right ones to ask and to subject to normative deliberation. This is precisely because emergent technologies are not yet embraced as practices and thus are not assignable to the stakeholders' outcomes (Vallor 2016). Practices are fixed modes of actions that aim at specific desirable values, such as to foster friendship or benevolence. Information technologies are still too dynamic and open and hence do not present options to act but different forms of life (Vallor 2016; Nida-Rümelin 1996, 2009). Vallor (2016) put it nicely when highlighting the fact that to ask whether "Is Twitter right or wrong?" (p.27) or even whether "Tweeting is right or wrong?" (p.27-28) are meaningless, empty questions, since they cannot take into account any particular existing contexts of shared actions. To make this point clearer: The rightness of an action cannot be determined by one single criterion or principle, the action's consequences or the agent's motivation alone. The desirability of the decision needs also to be evaluated against existing valued structures and norms. An ethical deliberation has to grasp questions such as: Does Twitter promote the value of "caring", "courtesy", "friendship"? Which values need to be taken into account and which practices need to be addressed within a social media platform that allows for short statements? How does Twitter respond to the demand of a respectful handling of each other despite diversity? Therefore, in the context of many ITs, it is not only necessary to ask about individual moral actions, but about the moral goodness of whole forms of life (cf. Nida-Rümelin 2005, 2009).

## 3 Embedding ethics into software development

Such ethical deliberations must be integrated into work processes in such a way that they are perceived as something positive. Individual software engineers need to be empowered and encouraged to reason normatively and be trained in ethical communication skills. We focus on software engineers because we address ethical issues that result from technical designs: Finding technical solutions for normative questions will not solve all ethical questions related to information technology. It will, however, promote more thoughtful designs and a responsible art of engineering in general. At the same time, it must be emphasized that those deliberations require structural embedding, but cannot be replaced by it, i.e., what is required is a process that enables and fosters ethical thinking. Therefore, we suggest including normative deliberation into development processes and thus make it part of documentation as well as quality assurance. Ideally, this approach links normative deliberation to the whole software lifecycle, for example by embedding agile ethics into DevOps methods, which integrate the development and the operation of software into a single methodology (Ebert et al. 2016). Such a proactive stance towards information technology is worked out e.g., by Brey (2000), Floridi (2008, 2011; cf. Russo 2010) and Coeckelbergh and Reijers (2020). They highlight, independent of their different theoretical foundation, the ability to influence real-life issues by changing technical design features. In this way, they expand the scope of action of the individual software engineer. A similar approach called Values in Design or Ethics by Design (cf. Simon 2016) has become popular since the 1990s. This ethical push emerged primarily in software engineering (van den Hoeven et al. 2015) to enable the deliberation of *embedded values*. The idea is that technology conveys values that have lifeworld implications, more precisely, that these values have moral consequences in the real world. Furthermore, these values implicitly embedded in technology reflect and perpetuate the attitudes and actions of its developers. Therefore, the focus throughout the entire development and deployment process is put on the project team's actions and decisions, which includes the software developer. Value-sensitive design (Friedman, Kahn and Borning 2002), ethics by design that primarily addresses AI issues (Richards and Dignum 2019), values in design (e.g. Manders-Huits 2009) or values in play (Flanagan, Howe and Nissenbaum 2008; Flanagan and Nissenbaum 2014) are similar methods to approach the embeddedness of values in digital technical products. They all oppose (strong) technological determinism and affirm individual autonomy to control technology and its outcomes. Hence, software engineering teams and companies have at least partial power and control over the design of their digital products, i.e., software engineering can intentionally exert influence on individuals (cf. van den Hoeven 2015). It is important to stress that incorporated values do have some sort of moral impact on real-life structures and are therefore subject to normative evaluation.

Embedded values approaches are a very practical enterprise that is intended to support the software developer in translating normative attitudes into technical objects. It can therefore be understood as a form of hermeneutics. Since the approach originates from software engineering, it is less theoretical or meta-theoretical, but brings with it a plethora of individual cases, which in themselves do not provide any real normative guidance (e.g., discussing one particular issue such as privacy by design or accountability). Thus, those approaches remain rather a second-degree decision aid, even in the case of value sensitive design that presents the most theoretical fundament (cf. Friedman and Hendry 2019). As values and normative principles are identified and prioritized, which is a theoretical and empirical endeavor, methods that integrate values in design will become necessary and helpful (cf. Grunwald 2015). However, little thought is given to the work environment or competences that are needed for such an approach to be successful (cf. Mittelstadt 2019).

Therefore, introducing ethical deliberation into the software development process from the very beginning is strongly recommended when building ethically sensitive machines and software. Of course, institutional support for ethical behavior is not only a question for software producers or developers (cf. Gogoll 2021). Yet, in comparison to other professional ethics such as medical ethics or business ethics, software engineering ethics needs to address peculiar demands in dealing with normative aspects simply because actions and decisions by software engineers are instantly translated into technical requirements and thus are so often perpetuated millionfold. Moreover, since software engineering lacks institutional sanctions and an overall professional commitment to consider public welfare (cf. Abbas et al. 2019), it is essential that the development team itself is capable of knowing how to guide itself through normative concerns. Hence, ethical deliberation and technical skills are competences that are required to achieve good normative design.

Thus, a humanistic, anticipatory ethics of information technology with regard to a professional group (software engineers) is needed, which includes elements of *disclosive ethics* (Brey 2001) and approves of a procedural character such as rational discursive moments (cf. Ott 2005). In order to understand such a humanistic approach as professional ethics, ethics must be understood as a guided practice that strengthens an *ethos*, i.e., habits that most professionals recognize as their purposes for action. For this, one needs a proactive ethics, i.e., the understanding that normative considerations must play a role from the very beginning and throughout the whole product lifecycle in order to be able to specifically develop normatively desirable artifacts. Since technology falls into the realm of possibility and uncertainty, and is thus intrinsically dynamic, we can only capture this through a processual ethics, i.e., rethinking the conditions, or to use Shilton's (2013) designation to introduce *value levers*, that make such a work ethic possible. This means that companies need to empower software engineering teams to undertake ethical deliberations, i.e., taking values, social goals and collective goods seriously in their considerations.

# 4 Implementing ethics in the agile software development environment

Although Talcott Parsons made the concept of agility part of organizational theory as early as the 1950s, it had only a limited influence on the 2001 *Agile Manifesto* set up by Kent Beck and other experienced software developers (Förster and Wendler 2012). In their manifesto they formulated four basic value slogans as the essence of agility (see http://agilemanifesto.org):

- Individuals and interactions over processes and tools,
- Working software over comprehensive documentation,
- Customer collaboration over contract negotiation,
- Responding to change over following a plan.

The Agile manifesto prioritizes individuals, working software, customer collaboration, and responding to change over a strict plan that needs to be followed, processes that need to be implemented, or documentation that needs to be filed. The goal of these principles is to provide simple responses to constantly changing external requirements and thereby avoiding costs generated by finalized but unsatisfactory software products – whether this is because of changing requirements, contexts, technological shortcomings or non-functional problems, which are often rooted in disregarding values. The objective formulated by agile methods is to produce working products while taking into account constantly changing environments. In agile processes, working (i.e., executable) products can continuously be presented to the customer who may test and evaluate the software immediately so that the current status quo of the product can serve as a basis for further deliberations (Flewelling 2018). The idea of agility is rooted in the belief that a software system can hardly be specified up front and that it will always be necessary to react to unexpected changes, because of unforeseen environmental impacts, technological development or simply because the customer changed their mind. Thus, agility can be defined as provided by the Advanced Research Programs Agency (ARPA) and the Agility Forum as "the ability to thrive in an environment of continuous and often unanticipated change" (Sarkis 2001, p. 88).

Due to its flexibility, agile processes have become the de facto standard for software development in non-regulated industries in recent years. According to a survey by digital.ai "95% of respondents report their organizations practice Agile development methods" (Digital.ai, 2020). The field of software engineering leading with about 37% of teams using the agile method to develop software. Arguably, respondents in these studies may overstate the level of agile adoption, e.g., how many teams are actually adhering to the complete toolkit of agile methods vs. only cherry-picking some aspects of it. Yet, it seems accurate that agility has "become a mainstream development methodology" (Shim and Lee, 2017) or even "the new normal" – especially when we take into account that many companies plan to adopt agile in the short and medium future (Koning and Koot, 2019; Hewlett Packard, 2017). Agile development methods such as Scrum, Extreme Programming or Kanban were used to anchor individual techniques such as user stories or pin boards in organizational cultures. By implementing them in work processes classic hierarchization (top-down approach) can be avoided in order to realize the values of agile programming. To analyze how ethical deliberations can be integrated in agile programming, we focus on one specific and also the most common (Digital.ai 2020) agile process, Scrum, as depicted in Figure 1. (cf. Schwaber and Beedle 2002; Cao and Ramesh 2008; Schwaber and Sutherland 2011).

As outlined above, the core idea of Scrum is that customer requirements and development conditions will change. The best way to satisfy a customer is to deliver small incremental releases that will slowly converge with the customer's wishes in fixed time intervals, so-called sprints that usually last 2-4 weeks. (It is not by chance that this incremental approach also alleviates a fundamental problem of software engineering, namely the integration of different subsystems.) A Scrum consists of a product owner, who interfaces with the stakeholders (mainly the customer and the users. However, this could also include representatives of the government, NGOs or other groups with an interest or stake in the final product (c.f. Sverrisdottir et al. (2014)) and represents their interests; developers who implement the functionality; and a scrum master who is responsible to remove any work impediments and ensures that the development process runs smoothly. While in traditional development processes requirements are often prioritized at the beginning of the project, agile projects often change the priority of requirements or even remove them in their entirety. In Scrum, the product owner will be part of every sprint planning meeting and jointly with the developers decide which requirements will be part of the next iteration. Requirements that are not selected for the next sprint remain in the so-called product backlog, essentially a list of requirements, and will, if they have not become obsolete, implemented in one of the next sprints. It is important to note that the sprint planning meeting is time boxed: for a two-week sprint it should last 4 hours and up to 8 hours for a four-week sprint. This means that the deliberations must come to an end and a decision must be reached. After this, if no major problems occur, the development team could focus on delivering code and not be stressed by making new decisions every day.

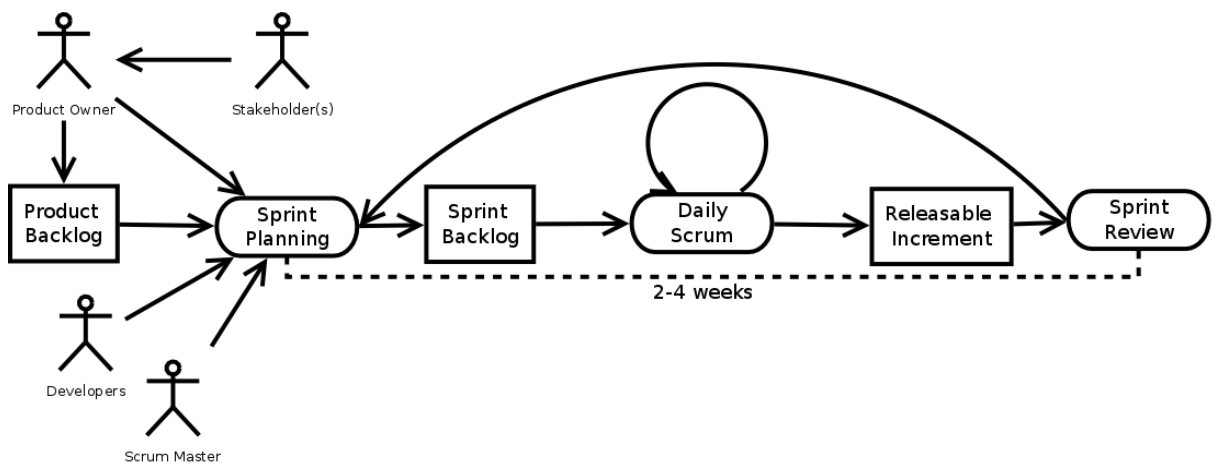

Figure 1: A schematic view of Scrum.

Of specific interest to our work is when and how requirements, which are normative in nature, are translated from being the goals set out by stakeholders to concrete requirements to be implemented by developers. It is precisely here where the scope of normative deliberation can be widened, and, consequently, values and attitudes must be addressed and made explicit. However, the *Agile Manifesto* (2001) did not address ethical issues when introducing its agile principles. It does, for instance, highlight the need for the customer's or the product owner's positive attitude towards changing requirements and fluid prioritization in exchange with the software team; but this does not explicitly include public interests (Judy 2009). This results in an incongruity in the formulation of technical, economic, and ethical requirements. Besides the integration of ethical deliberations in an agile work culture, it may also be important to foster moral attitudes such as whistle blowing or ethical peer pressure which are often neglected, after scandals such as Cambridge Analytica.

In the following, we will offer five reasons to illustrate why agile processes should be considered as a well-equipped platform to enable ethical deliberation during the development of software products.

## 4.1 Agile is already widely spread in the industry

As mentioned above, in order to avoid transition costs and minimize costs of implementation in general, we should look for ethical deliberation to be implemented in an already existing management system. While agile offers more than just a high rate of adoption, prevalence is one of the necessary conditions when trying to put theoretical considerations into actionable practice. If a method of developing ethically guided technical products required a fundamental change in a company's workflow, it is hard to imagine that it would succeed. The agile method being widespread in software engineering makes a symbiotic approach auspicious and more likely to have a positive effect on development in the near future. Agility being anchored in the software industry eases the introduction of ethical deliberation because it is integrated into a well-known, learned, and already working method of development. Consider, for instance, the fact that agile already has a well-defined process of meetings and planning. It seems obvious that the addition of ethical deliberation and the documentation of it will decrease the friction as opposed to the introduction of new processes.

Furthermore, since the specificity of a potential ethical issue depends on the concrete product and situation which makes a one-size-fits-all approach to tackle ethical issues during the development impossible, the use of an already implemented development regime offers the advantage of providing risk-free leeway for testing out various mechanisms to further ethical deliberation within the agile framework. For instance, a company could experiment with the introduction of ethical deliberation within teams as a follow-up to the daily sprint (a regular short meeting) every fortnight or add user stories (requirements) that deal with the experience of minorities, different perspectives etc. Thus, firms would be able to learn more about when and to what extent ethical deliberation is necessary depending on a project's context, its domain as well as team size, seniority of developers, and other relevant factors (cf. Spiekermann 2020). This could all be done with comparatively little additional effort for the developers since it would be implemented into the already existing framework of agile.

It is important to note that for agile frameworks to work, strong support of management is a necessity (Dikert et al. 2016). The same is true for ethical deliberation within the process. If ethics is perceived as a part of agile software development, hopefully, it can also piggyback on management's support of agile in general.

## 4.2 Flat hierarchies offer autonomy to developers and designers

As we have argued above, ethical deliberation requires that software developers have certain degrees of freedom in their work. In an environment in which developers strictly follow orders, work on a narrowly defined problem, and are thus isolated from interactions with other steps or actors in the development process, there is little room for ethical deliberation other than on the level of management. However, the information asymmetry that management faces especially in the IT domain, might render many ethical issues which might become obvious during the development process inaccessible to the decision makers at management level so that they remain unsolved. If we want to introduce moral attitudes within the development process and include developers in the act of detecting ethical pitfalls, incorporating stakeholder values, and developing normative desirable software, we must provide developers with greater freedom in the development process. Agile as a process offers just that. In fact, empowerment and autonomous decision making are seen as a key factor in regard to forming truly agile teams (Kidd, 1994; Van Oyen et al. 2001; Mudili, 2017).

If software developers and designers work in clearly defined hierarchies with a narrowly defined scope of action, there is little to no place for critical thinking. For example, employees who expect primarily monotonous work are more likely to accept organizational misconduct or attempt it themselves (cf. Staffelbach et al. 2014). In such contexts, critical thinking and the additional motivation required to attempt this kind of thinking are perceived as external factors and therefore something that can be delegated. Conversely, intrinsically motivated employees need a working environment that offers them varied tasks, enables them to have a good relationship with their colleagues and superiors, and provides them with a safe environment. This means that employees need more room of action so that they can feel autonomous and respected which in turn positively influences motivation (Noll et al. 2017; Law and Charron 2005; Melo et al. 2012). Furthermore, Stein & Untertrifaller (2020) found evidence in a laboratory experiment that "workers who prefer to work in an ethical work environment perform better if they are also responsible for it, compared to a situation where it was imposed on them" – underlining the effect of autonomy and responsibility on performance.

Autonomous persons are more likely to assume responsibility because they realize that they have the option to decide by themselves – to structure their life according to their well-justified choices.

Psychologically empowered employees strengthen agile teams because of their acknowledged cognitive abilities that underpin the relationship of the individual to its work such as meaningfulness, competence, self-determination, and impact (Breu et al. 2002; Muduli 2017). This aspect cannot be over-emphasized, because people must be able to integrate themselves into a social system according to their abilities. Therefore, employees must experience integration as cooperation and not as oppression (cf. Boes et al. 2020a, Boes et al. 2020b). Psychological empowerment in turn needs to be structurally integrated to enhance resilient individuals. This is exactly where agility comes into play. An illustrative example is the fact that the more supervised employees are, the less likely they are to trust other individuals (Grund and Harbring 2009). However, mutual trust allows for a more respectful and equal treatment of each other which in turn fosters a work culture that supports cooperative behavior and thus needs to rely less on contractual or institutionalized procedures (Mulki et al. 2006; Lester and Brower 2003; Wu et al. 2009).

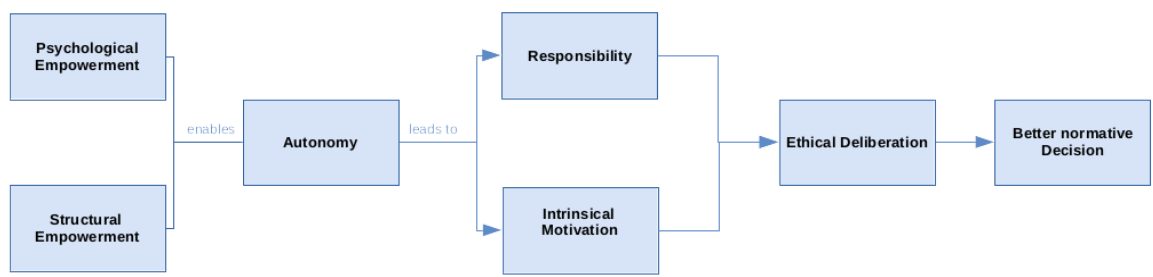

Figure 2: Empowerment, autonomy and ethical deliberation

Figure 2 shows the relationship between these concepts. If developers are structurally empowered their self-perception as respected autonomous beings and the feeling of responsibility increase their motivation (Valentine and Fleischman 2008; Koronios et al. 2019). This is a crucial cornerstone of ethical deliberation. In order to motivate people to look for ways to build normative desirable software products, we must provide the opportunity to deliberate, decide and execute the code based on that justified decision. In fact, if a software developer has no other choice but to implement software in a specifically predetermined way, we exclude the developer from the process of designing ethically sound software and have to rely on upper hierarchies to simply get the product normatively correct from the start. Hence, it will be more likely that ethical issues will only be recognized ex post. This, in turn, will either lead to them being ignored or require costly changes to alter the product after completion (Beck and Andres 2013, Ch. 13; Brey 2012).

## 4.3 Ethical deliberation needs functioning teams and participation

The third aspect why agile methods lend themselves to be enriched with ethical deliberation is the emphasis that is put on the value of teamwork and the constant sharing of knowledge. This is of great importance when we want to embed a procedure that is based on rationally discursive elements to formulate, prioritize, and decide upon values and their technical implementation. Additionally, the procedural approach will promote ethical habits within the software community, i.e., establish critical thinking and value explorations as normal every-day behavior with regard to designing and developing digital products. This is precisely what Vallor (2016) points to when she discusses virtues as the foundation of ethically sound software systems. Virtues form an *ethos* that stabilizes and structures good normative software engineering skills. This results in the creation of an environment in which developers not only think in terms of technical functionality but also in terms of ethically desirable features and products.

In agile processes, teams are given a high degree of autonomy, being referred to as "self-organized teams" in Scrum and "empowered teams" in extreme programming. As a best-case scenario, teams with high maturity and experience can independently organize their work and choose which resources should be applied to what part of the development task and at what time. An agile team needs to have the ability to understand and implement the items in the backlog which often leads to the fact that a team is to be composed of several people with different backgrounds and skills. The fact that the team has a certain degree of autonomy regarding the planning of the work requires constant communication between team members so that every single team member has to have at least a basic understanding of the skills of their teammates as well as the knowledge of what they are working on and how this integrates with their own work. Knowledge silos are thus less likely to emerge. Ethical deliberation also relies on communication among team members and the disruption of knowledge silos facilitates identifying ethical issues that emerge through the interaction of different parts of software. The obvious advantage in agile methods lies in the fact that teamwork is already at the core of the methodology and thus does not need to be artificially introduced, e.g., as a special ethics meeting between developers who otherwise do not share their work or have no personal connections. Interestingly, the strengthening of teamwork through the process of mutual discussions during ethical deliberations could also, in turn, provide a positive effect on the promotion of agility among developers (Mudili 2017).

## 4.4 Agile processes promote techno-ethical realism

Due to the iterative approach and the constant increments of software, ethical deliberations tend to be less abstract because reasoning is more concrete and object-focused: Since the goal of each sprint is a "working prototype" certain practical consequences of the design are already tangible during the development process. As outlined above, within the agile process incremental releases are continuously delivered so that unsatisfactory elements due to technological problems or in regard to usability can be detected early on and changed. Thus, many difficulties in handling and operation can be solved immediately. It also ensures that any extensive and rigid requirements set at the beginning of the project are constantly adjusted, weighed or, if necessary, dropped in order to be able to develop an adequate product. This can also be understood as a conservative stance within technology assessment methodologies – a stance that puts more focus on generic normative issues rather than on speculative ones (Brey 2012), i.e., big data refers to privacy issues: Due to the amount of data required, big data inevitably leads to ethical and legal deliberations in dealing with privacy, be it data storage and the corresponding questions of data security and accessibility or how this data has to be collected. Privacy is therefore a generic ethical issue. The *de facto* handling of existing software guides further thinking and specifies abstract ideas. Ethical deliberations benefit from this because normative concerns can be localized and responded to more directly through this step-by-step development. This, in turn, supports reasoning on normative questions at each level of abstraction without going astray, i.e., technological thinking about war drones does not entail general moral questions on the issue of a "just war" which is located at a political level. Normative expectations thus become tangible and less speculative, as they are aligned with concrete objects. Requirements that have not been met are therefore made visible, and requirements that do no longer seem reasonable because they leave the considered domain can be identified and rejected. In this sense, the ethical object-orientation emphasizes a techno-ethical realism.

## 4.5 Sprints offer a salient endpoint to deliberation

Another advantage that agile processes offer are the clearly structured time frames that are tied to specific deliverables ("working prototype at the end of the sprint"). Since normative deliberations, such as factual knowledge, can be inexhaustible, people tend to find themselves in a position in which they search for the perfect normative design or, if the deliberation is on an issue close to their heart, focus on the perfect implementation of every single detail. The psychological burden that one "has not done enough" might be an impediment that hinders the continuation of development.

While the statement that trade-offs matter in software design as they do in life is certainly trivial, it is still important to keep in mind that ethical deliberation is itself subject to empirical constraints. Among these are a realistic account of human (psychological) limitations and the economic costs of a (prolonged) ethical deliberation, namely in the form of search costs. It seems obvious that it is not sensible to spend a long time deliberating on small details with little effect just to aim for a maximization of the normative good design. Of course, the higher the stakes, the more time should be spent deliberating – but a final decision needs to be made eventually. Time and the associated costs are a limiting factor in every software development process and the time attributed to the deliberation of ethical issues should be spent wisely. Obviously, any deliberation that claims to produce any practical impact must (eventually) come to an end. Between the two extremes – an omniscient being who does not need to deliberate at all and imperfect beings with limited knowledge such as humans, who could deliberate endlessly – there is a sweet spot, the perfect balance between advantages of deliberating about how to implement normative features into software and the costs of prolonging deliberation (time to market etc.). In a way this mirrors the debate in psychology and economics between the two concepts of "maximizing" and "satisficing" which dates back to Simon (1956). If deliberation is a process "that begins in ignorance and ends in knowledge" (Johnson, 2007) and given the fact that a complete and maximizing ethical deliberation is not realistic, we must look for a reasonable "end point" at which the deliberation stops. The c*oncept of satisficing* calls this the "aspiration level." This level serves as a stopping rule once a certain level of "good enough" is achieved. Here, choosing the first alternative that satisfies an aspiration level, e.g., the considered value has been given proper and "good enough" consideration, instead of a continued deliberation process might be a more feasible solution.

Agile processes with their concrete focus on time frames offer clear endpoints at which deliberation must come to an end and a decision must be made: either stop deliberating when any issues seems sufficiently explored for the time being or, if deliberation needs to continue, actively (and intentionally) push it forward to the next sprint. Especially with regard to normative issues and the assumption of responsibility resulting from recorded and documented decisions, external guidelines help to reach a decision when it is due. In a sense this feature of agile processes lies within the domain of moral design (Gigerenzer 2010), which focuses on the creation of suitable environments for ethical deliberation to prosper rather than to focus on the values of any single developer.

## 5 Concluding Remarks

In this paper we argued that agile development processes can include and enhance ethical deliberation due to certain features of this particular organizational style – among them the focus on autonomy of the engineer, flat hierarchies, the existing focus on team collaboration, the provision of techno-ethical realism, and the introduction of clear endpoints for ethical deliberation.

While it is a promising idea that a widespread development style can foster ethical deliberation during the development process, it is also necessary to think about how, when, and who must tackle these questions within management. It is also necessary to think about the possibility of formalizing ethical procedures regarding generic as well structural ethical topics and implement them within the agile framework. Furthermore, ethically informed engineering has to entail continuous observation and possibly alteration of digital artifacts throughout their entire life cycle focusing on normative good design via DevOps.